\documentclass[aps,prb,twocolumn,groupedaddress,showkeys,showpacs]{revtex4-1}

\usepackage{graphicx}
\usepackage{amsmath}
\usepackage{amsfonts}
\usepackage{footnote}
\usepackage{gensymb}
\usepackage{color}

\begin{document}

\title{Tamm-Langmuir surface waves}

\author{K.\,U.~Golenitskii}
\author{K.\,L.~Koshelev}
\author{A.\,A.~Bogdanov}
\email[]{bogdan.taurus@gmail.com}
\affiliation{ITMO University, 197101 St.-Petersburg, Russia}
\affiliation{Ioffe Institute, 194021, St.-Petersburg, Russia}

\date{November 11, 2013}

\begin{abstract}

In this work we develop a theory of surface electromagnetic waves localized at the interface of periodic metal-dielectric structures. We have shown that the anisotropy of plasma frequency in metal layers lifts the degeneracy of plasma oscillations and opens a series of photonic band gaps. This results in appearance of surface waves with singular density of states -- we refer to them as Tamm-Langmuir waves. Such naming is natural since we have found that their properties are very similar to the properties of both bulk Langmuir and surface Tamm waves. Depending on the anisotropy parameters, Tamm-Langmuir waves can be either forward or backward waves. Singular density of states and high sensitivity of the dispersion to the anisotropy of the structure makes Tamm-Langmuir waves very promising for potential applications in nanophotonics and biosensing.        

%
%
%
%
%
%
%

\end{abstract}

\pacs{42.79.Gn, 73.20.Mf, 73.40.Rw, 78.20.Bh}
\keywords{Surface plasmon polariton, bulk plasmon, anisotropy, metal-dielectric structure, surface waves}

\maketitle


\section{Introduction}

Surface electromagnetic waves can propagate along an interface separating two media while being localized in the transverse direction. There are many types of surface waves including surface plasmon polariton (SPP), optical Tamm waves, D'yakonov waves, Bloch surface waves etc.\cite{Polo2011,Maier2005,Pitarke2007,vinogradov2010surface,Iorsh2011,Yermakov2015,Xiang2014,Zapata-Rodriguez2013,Dyakonov1988}  Localization of surface waves can reach deep subwavelength level of up to 10 nm at the near-infrared range.\cite{choo2012nanofocusing} High localization is accompanied by strong amplification of evanescent electromagnetic fields. Such functionality makes surface waves relevant for many 	applications from subwavelength focusing,\cite{yin2005subwavelength,bartal2009subwavelength} imaging\cite{barnes2003surface,fang2005sub} and high-resolution lithography\cite{luo2004surface} to on-chip signal processing,\cite{tanaka2003simulations,Nikolajsen2004,zia2006plasmonics} non-linear optics, sensing, high-performance absorbers,\cite{wu2011large, Chen2016}, photovoltaics,\cite{atwater2010plasmonics} optical trapping\cite{Min2013, Zhao2016} and manipulation.\cite{Rodriguez-Fortuno2015,Petrov2016}

For sensing and spectroscopy applications is often needed to modify the emission rate (enhance or inhibit) of emitters (quantum dots, single atoms or molecules, NV-centers etc) placed in the vicinity of a substrate. In this case, the role of surface optical states (surface waves) becomes of the utmost importance. In particular, a modification of spontaneous emission rate due to guided surface mode of isotropic metasurface was analyzed in Ref.~[\onlinecite{Lunnemann2016}]. In Refs.~[\onlinecite{Neogi2002, Chang2006,PhysRevB.84.073403}] it was shown that SPP can substantially increase the rate of spontaneous emission.  However, enhancement of spontaneous emission due to SPP has a resonant behavior since density of optical states tends to infinity at the single frequency -- the frequency of surface plasmon resonance, where the group velocity of SPP tends to zero. To achieve the broadband enhancement we need surface waves with high density of states in a broad frequency range.

In this work we develop a theory of surface electromagnetic waves with broadband singular density of states, which are supported by periodic metal-dielectric structures with anisotropic conducting layers.\footnote{Under the term ``metal'' we understand any medium with free carriers.  It can be plasma, real metal, superconductor, doped semiconductor etc.}   In the literature structures consisting of alternating layers of isotropic dielectric and hyperbolic metamaterials are called ``multi-scale hyperbolic metamaterial" or ``photonic hypercrystal".\cite{Golenickij2013,zhukovsky2014photonic,smolyaninova2014self,huang2014veselago,narimanov2015dirac,PhysRevX.4.041014} We show that the broadband singularity of density of states comes as a result of infinite density of bulk plasmons (Langmuir modes) localized in the metallic layers. The anisotropy of these layers lifts the degeneracy of plasma oscillations and opens a series of photonic bands and gaps. Localized surface states analyzed in this work appear in the opened photonic gaps where  anisotropic layers exhibit the properties of a hyperbolic medium.\cite{Poddubny2013} We show that the surface waves under consideration inherit the main properties of bulk Langmuir modes.

The paper is organized as follows. 
In Sec.~\ref{sec:model} we introduce our model and main equations. 
The analysis of dispersion, field distribution and dissipation of Tamm-Langmuir surface waves is represented in Secs.~\ref{sec:dispersion} and~\ref{sec:dissipation}.
In Sec.~\ref{sec:reflection} , we propose the scheme of an experiment for the detection of the Tamm-Langmuir surface waves and model it numerically. The results of the work are summarized in Sec.~\ref{sec:conclusion}.  
  
\section{model}
\label{sec:model}
\subsection{Geometry of structure}
\begin{figure}
	\includegraphics[width=0.95\columnwidth]{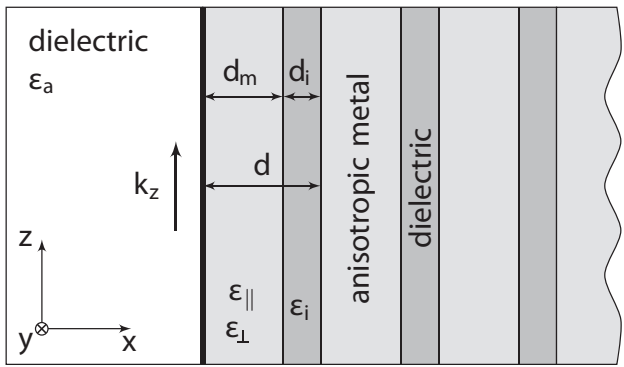}
	\caption{Semi-infinite periodic metal-dielectric structure with anisotropic conducting layers.}
	\label{fig:model}
\end{figure}

Let us consider a semi-infinite periodic layered structure with a period $d$ (Fig. \ref{fig:model}). 
We suppose that the unit cell of the structure consists of two layers. 
The first one is a dielectric of thickness $d_i$ with constant permittivity $\varepsilon_i$. 
The second one is an anisotropic conducting material with thickness $d_m$ and dielectric function
\begin{equation}
   \varepsilon_m=
   \left(
   \begin{matrix} 
      \varepsilon_{\bot}(\omega) & 0 & 0\\
      0 & \varepsilon_{||}(\omega) & 0\\
      0 & 0 & \varepsilon_{||}(\omega) \\
   \end{matrix}
   \right). 
\end{equation}
We suppose that each tensor component depends on the frequency within Drude-Lorentz approximation:
\begin{equation}
\varepsilon_s=\varepsilon_s^\infty\left(1-\frac{\Omega_{s}^2}{\omega(\omega+i\gamma_s)}\right), \ \ \ \ \ \ \ \ s=\bot, ||.
\label{Drude_epsilon}
\end{equation}
Here, $\Omega_{\bot, ||}$ are the plasma frequencies across and along the layers, $\gamma_{\bot, ||}$ are the damping frequencies. Therefore, optical axis of anisotropic layers perpendicular to the interface of the structure. Henceforth, we suppose that $\varepsilon_\bot^\infty=\varepsilon_\|^\infty=\varepsilon^\infty$.

Some natural materials (bismuth,\cite{PhysRev.120.1943} graphite,\cite{Sun2011} h-BN \cite{brar2014hybrid}) possess anisotropic plasma frequency. However, the anisotropy can also be induced artificially.  
For example, in plasma it is achievable via imposing the external magnetic field.\cite{bernstein1958waves} 
In bulk semiconductors it can be accomplished by the implementation of superlattice.\cite{Grecu1973,Koshelev2015,Sun2011,Bogdanov2011,Bogdanov2012}  
Alternatively, in wire medium, which dielectric function in the GHz frequency region can be described within the Drude approximation, the anisotropy of plasma frequency can be reached due to the difference in the periods along the proper directions.\cite{Belov2003,Silveirinha2003}
The results represented below are scalable and can be applied for different metamaterials operating from the visible to terahertz and radio frequencies (whenever the Drude model is justified).


\subsection{Dispersive equation}


Let us derive the dispersion equation for surface wave propagating along the interface between a periodic medium and a semiinfinite dielectric cladding. 
We will seek the solutions of the Maxwell's equations in the form of TM-polarized wave traveling along the $z$-direction. 
This means that all field components depend on $z$ and~$t$ as $\exp(ik_zz-i\omega t)$.   
The electromagnetic field of the surface waves should tend to zero away from the interface, while on the interface the tangential components of electric and magnetic field should be continuous. This condition yields for TM-polarized waves the following dispersion equation:\cite{yeh1988optical} 
\begin{equation}
\frac{\exp(ik_bd)-M_{11}}{M_{12}}=\frac{ck_x}{\varepsilon_{\rm a}\omega}.
\label{disp_eq_1}
\end{equation}
Here, $k_x^2=\frac{\varepsilon_{\rm a}\omega^2}{c^2}-k_z^2$, $k_b$ is the Bloch wave vector, $\varepsilon_{\rm a}$ is the permittivity of semiinfinite cladding layer, $M_{ij}$ is a component of the transfer matrix corresponding to single period of the multilayered structure.   \cite{yeh1988optical} 
Bloch wave vector $k_b$ in Eq. (\ref{disp_eq_1}) obeys the dispersion equation for periodic structure
\begin{equation}
2\cos(k_bd)=M_{11}+M_{22}.
\label{disp_eq_2}
\end{equation}
Equations (\ref{disp_eq_1}) and  (\ref{disp_eq_2}) define the dispersion of the surface waves $\omega(k_z)$. 
The sign of the imaginary parts of wave vector component $k_x$ and Bloch wave vector $k_b$ should be chosen  from the condition that electromagnetic field of surface waves tends to zero away from the interface.

In what follows we will use dimensionless quantities:
\begin{equation}
   \begin{gathered}
   \widetilde{k}_z=k_z d; \quad \widetilde{\omega}=\frac{\omega d}{c}; \quad \widetilde{\gamma} = \frac{\gamma d}{c}\\ 
   \widetilde{x} = \frac{x}{d}; \quad \xi=\frac{d_m}{d}.
   \end{gathered}
\end{equation}
This means that we measure spatial scale in units of the period. 
 
\section{Results of calculations} 
\label{sec:dispersion}
\subsection{Isotropic metal-dielectric structure}
\begin{figure*}
\includegraphics[width=\linewidth]{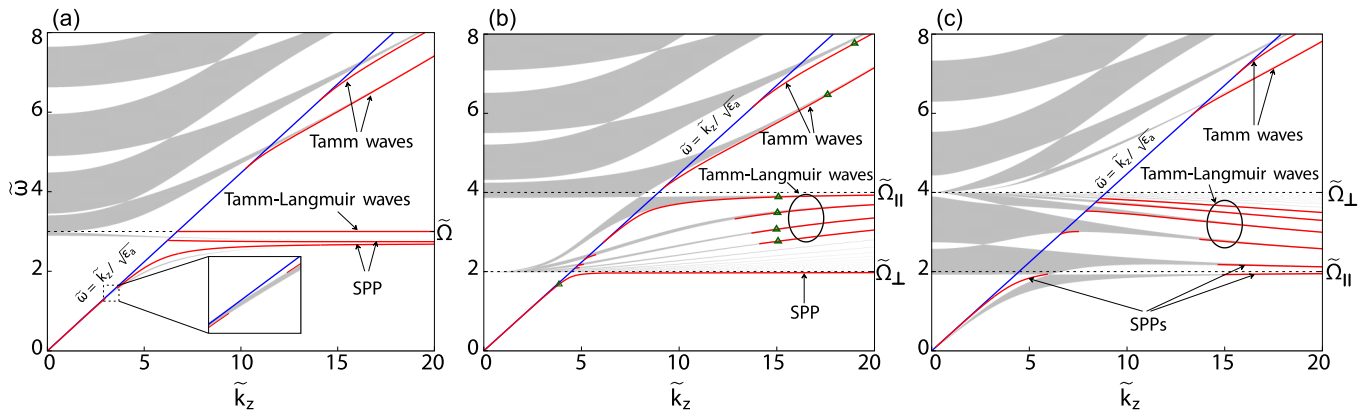}
\caption{Dependence of $\omega$ on $k_z$ for surface waves on the interface of metal-dielectric structure (Fig.~\ref{fig:model}). Gray areas correspond to the allowed band of the structure. In the case (a) $\widetilde{\Omega}_{\perp}=\widetilde{\Omega}_{||}=\widetilde{\Omega}=3$, (b) $\widetilde{\Omega}_{\perp}=2$ and $\widetilde{\Omega}_{||}=4$ (c) $\widetilde{\Omega}_{\perp}=4$ and $\widetilde{\Omega}_{||}=2$. Other structure parameters are common for (a), (b), (c): $\xi=0.87$,  $\varepsilon^\infty_m=9.5$, $\varepsilon_i=1.8$, $\varepsilon_{\rm a} = 5$. Electric field distribution for the regions marked by green triangles is presented in Fig.~\ref{fig:fields}.}
\label{fig:dispersion}
\end{figure*}

\begin{figure}[htbp]
   \centering
   \includegraphics[width=0.9\columnwidth]{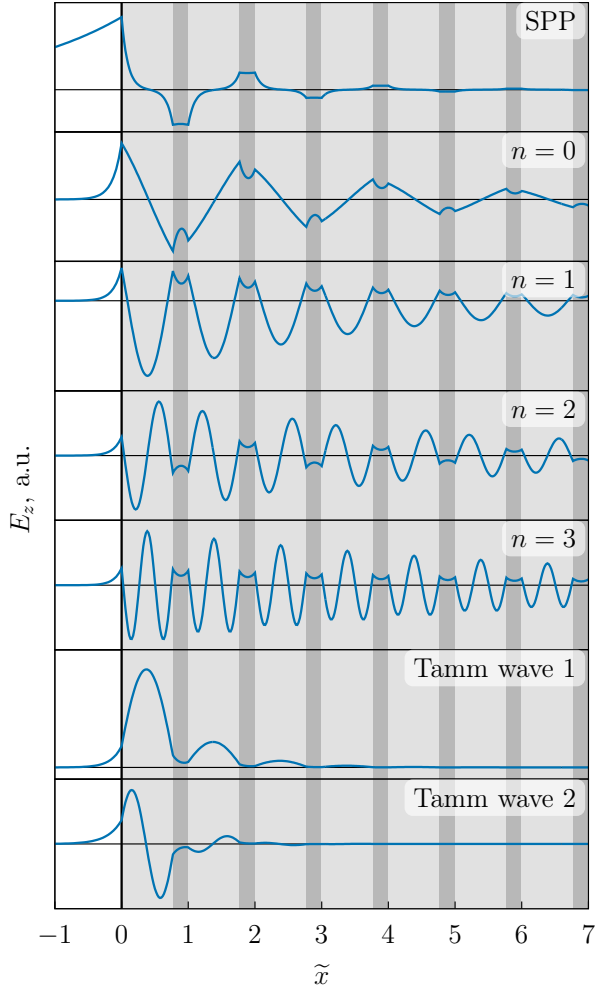} 
   \caption{Field component $E_z$ distribution for surface waves on the interface of anisotropic metal-dielectric structure. Parameters are $\widetilde{\Omega}_{\perp}=2$, $\widetilde{\Omega}_{||}=4$, $\xi=0.77$,  $\varepsilon^\infty_m=9.5$, $\varepsilon_i=1.8$, $\varepsilon_{\rm a} = 5$. Presented modes correspond to green triangles on dispersion curves in Fig.~\ref{fig:dispersion}(b) arranged in ascending of frequency.}
   \label{fig:fields}
\end{figure}

At first, let us analyze the spectrum of the surface waves localized at the interface of an isotropic metal-dielectric structure neglecting the dissipation. This means that $\widetilde{\Omega}_{\|}=\widetilde{\Omega}_{\bot}=\widetilde{\Omega}$ and $\widetilde{\gamma}_{\|,\bot}=0$ in Eq.~(\ref{Drude_epsilon}). 
The dispersion of surface waves for this case is shown in Fig.~\ref{fig:dispersion}(a). 
Dimensionless parameters were taken as follows: $\xi=0.77$,  $\varepsilon_m=9.5$, $\varepsilon_i=1.8$, $\widetilde{\Omega}=3$, $\varepsilon_{\rm a} = 5$. 
Gray areas correspond to the allowed photonic bands of the structure. 
One can see that the spectrum contains two types of surface waves. 
The waves of the first kind are conventional Tamm waves.
Their properties have already been analyzed in detail elsewhere (see, for example, Ref.~[\onlinecite{Bulgakov2004}]). 
The frequencies of Tamm waves are above the plasma frequency where metal layers exhibit dielectric behaviour. 
Therefore, metal-dielectric structure in this case can be considered as a dielectric Bragg reflector. 
The distribution of electromagnetic field for Tamm waves is shown in Fig.~\ref{fig:fields}. 
One can see that electromagnetic field oscillates inside the layers and exponentially decreases away from the interface. 

Second type of surface wave which can propagate along the interface of isotropic metal-dielectric structure is surface plasmon polariton. Its properties are well-documented in different structures (see, e.g., Ref.~[\onlinecite{Pitarke2007}]). in Fig.~\ref{fig:dispersion}(a) we can see two SPP modes: symmetrical and asymmetrical.

The line $\widetilde\omega=\widetilde\Omega$ corresponds to bulk plasmons (Langmuir waves).  
Langmuir wave represents collective oscillations of free charges. 
If all charges are identical and do not interact with each other then we have set of unbound oscillators with same eigenfrequency $\widetilde\Omega$. 
In this particular case Langmuir modes are actually not waves since their group velocity is zero and they do not transfer energy. 
If we take into account the interaction between oscillators it results in the appearance of dispersion and non-zero group velocity.\cite{Rukhadze1961,Forstmann1986}

Magnetic field of the Langmuir waves is equal to zero, i.e. these are pure electric waves. 
The electric field is perfectly confined inside the metal layers. 
Therefore, metal-dielectric structure represents a set of unbound metal waveguides. 
Electric field of Langmuir waves can have arbitrary spatial distribution inside the layers but should be equal to zero on the boundaries of the layers. This means that Langmuir modes are infinitely degenerated. 
However, in reality the degeneration degree is finite due to the spatial dispersion which is generally exist.


\subsection{Anisotropic metal-dielectric structure}     

 Let us now analyze the spectrum of the surface waves localized at the interface of an anisotropic metal-dielectric structure neglecting dissipation ($\widetilde\gamma=0$). Two cases are possible: (i) $\widetilde\Omega_{\bot}<\widetilde\Omega_{||}$; (ii) $\widetilde\Omega_{\bot}>\widetilde\Omega_{||}$. 
Dispersion of the surface waves for both of these cases is shown in Figs.~\ref{fig:dispersion}(b) and~\ref{fig:dispersion}(c). 
One can see that the degeneracy of Langmuir modes is lifted and the line $\widetilde\omega=\widetilde\Omega$ splits into a set of allowed bands corresponding to different Langmuir modes. 
All of these bands are squeezed between $\widetilde\Omega_{\bot}$ and $\widetilde\Omega_{||}$, thus their density of states remains singular.
In each stop band there is one state corresponding to the surface wave but in Fig.~2 we show only first four modes.  Frequency region of the singularity is determined by anisotropy, i.e. by plasma frequencies $\widetilde\Omega_{\bot}$ and $\widetilde\Omega_{||}$. It is worth emphasizing that these modes exist alongside with SPP and surface Tamm waves.

The dispersion for the surface wave under consideration is positive if $\widetilde\Omega_{\perp}<\widetilde\Omega_{||}$ and negative if $\widetilde\Omega_{\perp}>\widetilde\Omega_{||}$. Therefore, these waves
can be either forward or backward waves.
The same conditions are fulfilled for Langmuir modes in plasma waveguide.
The field distribution for these surface waves [see Fig.~\ref{fig:fields}] is similar to the Langmuir modes in the plasma waveguide --
electromagnetic field oscillates in the conducting layers and have exponential behavior in dielectric layers.\cite{Bogdanov2012} 
It might be asserted that the surface wave under consideration inherits the main properties of Langmuir waves. 
Therefore, it is quite natural to call them {\sl Tamm-Langmuir waves}.
An important feature is that a single Tamm-Langmuir state always exists between allowed bands corresponding to Langmuir modes. 

As we mentioned above, the number of Tamm-Langmuir waves is infinite in our model.
In a real system, their number is finite at least due to spatial dispersion at short wavelengths.\cite{koshelev2016interplay}

\section{Dissipation of Tamm-Langmuir waves}
\label{sec:dissipation}
To take into account losses in metallic layer we put $\widetilde\gamma_{\perp,\|} \neq 0$ in Eq.~\eqref{Drude_epsilon}. For the sake of simplicity losses are supposed to be isotropic, i.e. $\widetilde{\gamma}_\| = \widetilde{\gamma}_\perp = \widetilde{\gamma}$. Figure of merit (FOM) for the surface waves can be introduced as
\begin{figure}
	\centering
	\includegraphics[width=\columnwidth]{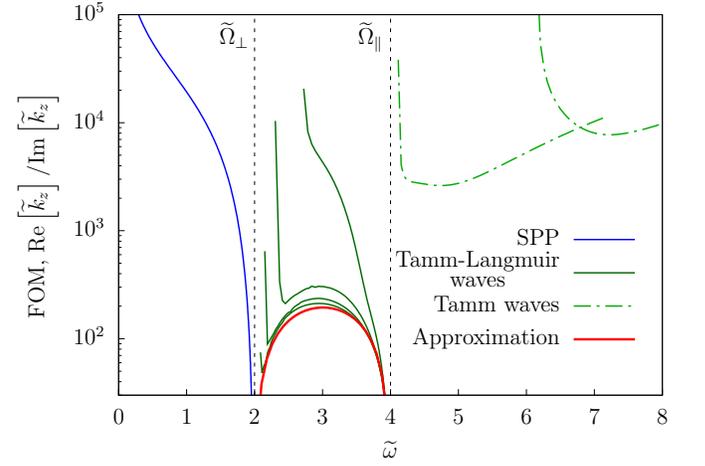} 
	\caption{Figure of merit for surface waves in case of positive dispersion Fig.~\ref{fig:dispersion}(b). Analytical approximation for Tamm-Langmuir waves is given by Eq.~\ref{eq:FOM}. $\widetilde{\gamma}_\| = \widetilde{\gamma}_\perp = \widetilde{\gamma} = 0.01$.}
	\label{fig:fom}
\end{figure} 
\begin{equation}
\mathrm{FOM}\,(\widetilde\omega) = \frac{\mathrm{Re}[\widetilde k_z]}{\mathrm{Im}[\widetilde k_z]}.
\end{equation}
The meaning of such defined FOM is the free path measured in the wavelength units. So, higher FOM corresponds to smaller losses and bigger propagation length. Figure~\ref{fig:fom} shows FOM for surface electromagnetic waves for $\widetilde{\gamma}=0.01$. In the vicinity of the light line ($\widetilde{\omega}=\widetilde{k}_z/\sqrt{\varepsilon_{\rm a}}$) the Tamm-Langmuir modes are weakly localized and their FOM tends to infinity. The same situation is observed for Tamm waves near the cutoffs and for SPP at low frequencies.  

Drop of FOM for the Tamm-Langmuir waves near $\widetilde\Omega_\|$ or $\widetilde\Omega_\bot$ is explained by the fact that the modes are mainly concentrated in the metal layers where absorption occurs. Analytical expression for the spectrum of FOM can be derived from Eq.~(\ref{disp_eq_1}) under assumption that $\mathrm{Re}[\varepsilon_{\|,\perp}] \gg \mathrm{Im}[\varepsilon_{\|,\perp}]$:
\begin{equation}
\label{eq:FOM}
\mathrm{FOM}\,(\widetilde{\omega}) =\frac{2}{\widetilde\gamma\widetilde\omega}\left|\frac{(\widetilde\omega^2-\widetilde\Omega_\perp^2)(\widetilde\omega^2-\widetilde\Omega_\|^2)}{\widetilde\Omega_\perp^2 - \widetilde\Omega_\|^2}\right|.
\end{equation}
Here, $\widetilde\omega$ is supposed to be between $\widetilde{\Omega}_\perp$ and $\widetilde{\Omega}_\|$. Eq.~\eqref{eq:FOM} is plotted in Fig.~\ref{fig:fom} with red solid line. One can see that it is in good agreement with numerical results for the modes with $n\gg1$.

\begin{figure}
	\centering
	\includegraphics[width=\columnwidth]{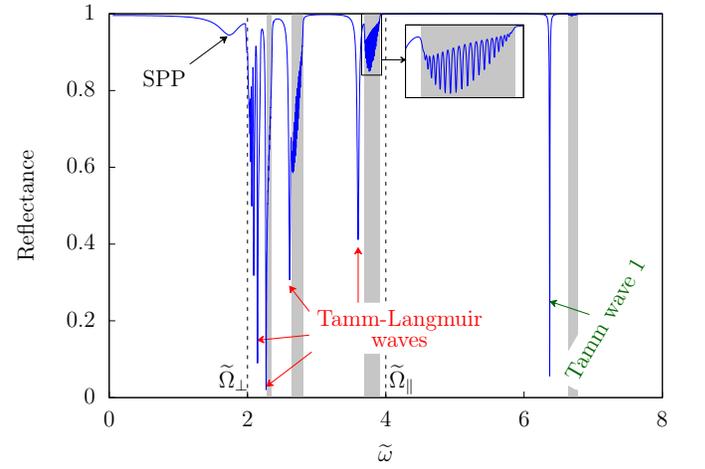} 
	\caption{Frequency dependence of reflection coefficient for incident angle of $45^\circ$. Gray color corresponds to allowed bands in infinite periodic structure.}
	\label{fig:reflection_section}
\end{figure}

\section{Simulation of experiment}
\label{sec:reflection}
\begin{figure}
	\centering
	\includegraphics[width=1.0\linewidth]{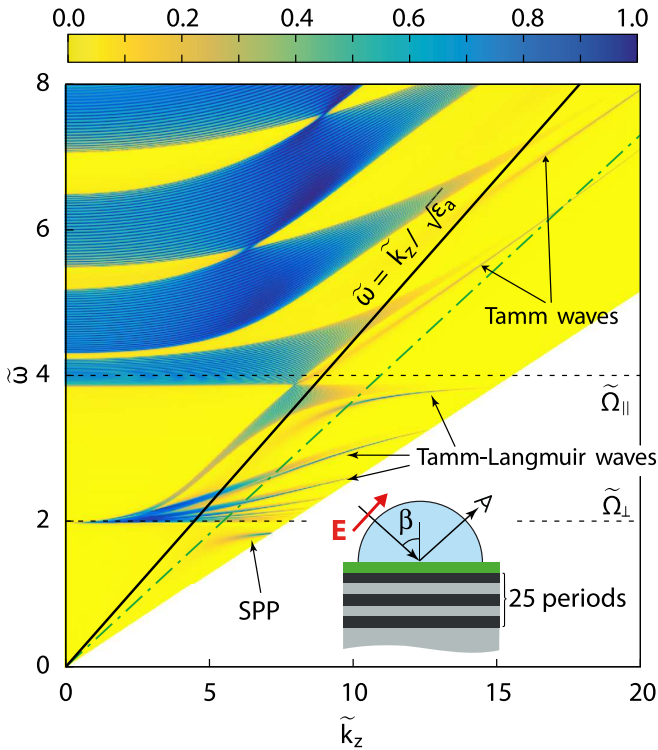} 
	\caption{Reflection coefficient map from finite structure with number of periods $N=25$.  A cut-off of reflection coefficient for incident angle of $45^\circ$ is shown with green dashed line (see Fig.~\ref{fig:reflection_section}). }
	\label{fig:reflection}
\end{figure}

Spectrum of Tamm-Langmuir modes can be analyzed in the experiment where a plane TM-polarized electromagnetic wave impinges onto the interface of a periodic metal-dielectric structure at an angle $\beta$ via a hemispherical high-index dielectric prism coupled to the sample through a thin low-index dielectric layer (Otto configuration).\cite{Otto1968} The scheme of the experiment is shown in the inset of Fig.~\ref{fig:reflection}. 

As an example, we consider a structure with the parameters from Fig. \ref{fig:dispersion}(b). We suppose that the structure consists of 25 periods and is placed on a substrate with permittivity $\varepsilon_i$. The prism permittivity $\varepsilon_p$ is supposed to be equal to 15. The thickness and the permittivity of a dielectric layer separating the prism from the sample are $d_m/2$ and $\varepsilon_{\rm a}$, respectively.
 


The reflectance spectrum for the incident angle $\beta=45\degree$ is shown in Fig. \ref{fig:reflection_section}. Gray areas mark the allowed photonic bands. The minima in the photonic bandgaps correspond to the excitation of surface waves. Wide dip at low frequencies ($\widetilde{\omega}<\widetilde{\Omega}_\bot$) corresponds to SPP whereas the dip at high frequencies ($\widetilde{\omega}>\widetilde{\Omega}_\bot$) is related to the Tamm wave. The series of dips at $\widetilde{\Omega}_\bot<\widetilde{\omega}<\widetilde{\Omega}_\|$ is a manifestation of Tamm-Langmuir modes.


A more full and illustrative picture can be obtained by measuring of the reflection coefficient map -- its dependence on both incident angle $\beta$ and frequency $\widetilde{\omega}$ (Fig.~\ref{fig:reflection}). The comparison of Figs.~\ref{fig:reflection} and \ref{fig:dispersion}(b) shows that the reflection coefficient map completely reproduces the numerically calculated spectrum of surface states. The reflection spectrum shown in Fig.~\ref{fig:reflection_section} corresponds to a section of the reflection coefficient map along the green dot-dashed line. Blue areas correspond to the allowed photonic bands. Since the multilayered structure is a slab with finite thickness, the allowed photonic bands have fine structure. Each of them consists of 25 (number of the periods) tight resonances corresponding to the waveguide modes of the slab.



\section{Conclusions}
\label{sec:conclusion}

We have shown that the anisotropy of plasma frequency in a periodic metal-dielectric structures lifts the degeneracy of plasma oscillations in the metal layers and opens a series of photonic band gaps. This results in the appearance of surface waves with singular density of states -- Tamm-Langmuir waves. Such naming is natural since: (i) their dispersion is very similar to the one of bulk Langmuir modes (bulk plasmons) in hyperbolic metamaterial waveguides (ii) they belong to the photonic band gaps just like Tamm waves.  Tamm-Langmuir waves coexist with SPPs and surface Tamm waves. Depending on the anisotropy parameters, Tamm-Langmuir waves can be either forward or backward waves. High density of optical states and low group velocity makes Tamm-Langmuir waves very promising for many applications from nanophotonics to biosensing.   

\acknowledgments

This work has been supported by RFBR (16-37-60064, 15-32-20665), by the President of Russian Federation (MK-6462.2016.2), and by program of Fundamental Research in Nanotechnology and Nanomaterials of the Russian Academy of Science. Numerical simulations have been supported by the Russian Science Foundation (Grant \#15-12-20028). The authors are grateful to I.\,S.~Sinev and A.\,K.~Samusev for useful discussions.
 
\bibliography{PaperBib}
\end{document}